\documentclass[11pt]{article}
\usepackage{graphicx}

\newcommand{\BABARPubYear}    {04}

\newcommand{\BABARConfNumber} {008}
\newcommand{\SLACPubNumber} {10660}

\input pubboard/babarsym
 
\newcommand{\ppmm}[2]{\penalty900\mkern5mu^{+\penalty900\mkern5mu{#1}}_{-\penalty900\mkern5mu{#2}}}

\newcommand{\dspace}{\ensuremath{\phantom{0}}}

\def\figurebox#1#2#3{%
    \def\arg{#3}%
    \ifx\arg\empty
    {\hfill\vbox{\hsize#2\hrule\hbox to #2{\vrule\hfill\vbox to #1{\hsize#2\vfill}\vrule}\hrule}\hfill}%
    \else
    {\hfill\epsfbox{#3}\hfill}%
    \fi}

\long\def\inst#1{\par\nobreak\kern 4pt\nobreak
    {\it #1}\par\vskip 10pt plus 3pt minus 3pt}

\begin{document}
{\pagestyle{empty}

\begin{flushright}
\babar-CONF-\BABARPubYear/\BABARConfNumber \\
%\babar-PUB-\BABARPubYear/\BABARPubNumber \\
SLAC-PUB-\SLACPubNumber \\
%hep-ex/\LANLNumber \\
August, 2004 \\
\end{flushright}

\par\vskip 5cm

% Title of the paper
\begin{center}
{\Large \bf \boldmath
Measurement of the Branching Ratio $\Gamma(\Dss\to\Ds\piz)/\Gamma(\Dss\to\Ds\g)$}

\end{center}
\bigskip

\begin{center}
\large The \babar\ Collaboration\\
\mbox{ }\\
\today
\end{center}
\bigskip \bigskip

% Abstract
\begin{center}
\large \bf Abstract
\end{center}
%To measure the branching fraction of the isospin violating decay $\Dss\to\Ds\piz$,
%the decays $\Dss\to\Ds\piz$ and $\Dss\to\Ds\g$ have been reconstructed using $90.4\invfb$
%of data recorded by the \babar\ detector at the \pep2\ \epem\ storage ring.
%%The reconstruction
%%efficiencies have been determined with signal Monte Carlo events; a measurement of the ratio
%%$\Gamma(\Dstarz\to\Dz\piz)/\Gamma(\Dstarz\to\Dz\g)$ has been used to confirm the correctness
%%of the method.
%Additionally, the decays $\Dstarz\to\Dz\piz$ and $\Dstarz\to\Dz\g$ have been reconstructed
%using the same data set.
%Our preliminary results for the branching ratios are
%$\Gamma(\Dss\to\Ds\piz)/\Gamma(\Dss\to\Ds\g) = 0.0621\pm0.0049\stat\pm0.0063\syst$ and
%$\Gamma(\Dstarz\to\Dz\piz)/\Gamma(\Dstarz\to\Dz\g) = 1.740\pm0.020\stat\pm0.125\syst$.
%
The ($\Delta I = 1$) decay $\Dss\to\Ds\piz$ and
($\Delta I = 0$) decay $\Dss\to\Ds\g$ have 
been reconstructed using $90.4$~$\invfb$ of data recorded by the \babar\  
detector at the \pep2\ asymmetric-energy $e^+e^-$ collider. The relative 
branching 
fraction of isospin violating decay to isospin conserving decay has 
been measured. Our preliminary result for this ratio is 
$\Gamma(\Dss\to\Ds\piz)/\Gamma(\Dss\to\Ds\g) = 0.0621\pm0.0049\:\stat\pm0.0063\:\syst$.
In addition we reconstruct the decays
$\Dstarz\to\Dz\piz$ and $\Dstarz\to\Dz\g$ for completeness and measure a 
preliminary relative branching fraction to be 
$\Gamma(\Dstarz\to\Dz\piz)/\Gamma(\Dstarz\to\Dz\g) = 1.740\pm0.020\:\stat\pm0.125\:\syst$.
Both measurements represent significant
improvements over present world averages.

\vfill
\begin{center}

Submitted to the 32$^{\rm nd}$ International Conference on High-Energy Physics, ICHEP 04,\\
16 August---22 August 2004, Beijing, China

\end{center}

\vspace{1.0cm}
\begin{center}
{\em Stanford Linear Accelerator Center, Stanford University, 
Stanford, CA 94309} \\ \vspace{0.1cm}\hrule\vspace{0.1cm}
Work supported in part by Department of Energy contract DE-AC03-76SF00515.
\end{center}

\newpage
} % end of pagestyle{empty}

\begin{center}
\small

The \babar\ Collaboration,
\bigskip

%% author list as of 02-Jul-2004 (609 authors)
%
B.~Aubert,
R.~Barate,
D.~Boutigny,
F.~Couderc,
J.-M.~Gaillard,
A.~Hicheur,
Y.~Karyotakis,
J.~P.~Lees,
V.~Tisserand,
A.~Zghiche
\inst{Laboratoire de Physique des Particules, F-74941 Annecy-le-Vieux, France }
% ...

\end{center}\newpage

The hadronic decay%
\footnote{Inclusion of the charge conjugated state is assumed throughout this paper.}
$\Dss\to\Ds\piz$ %~\cite{footnote:ChargeConj}
violates isospin
conservation and should be suppressed with respect to the
radiative decay $\Dss\to\Ds\g$. Cho and Wise~\cite{bib:ChoWise}
have calculated the decay
rate for $\Dss\to\Ds\piz$ using chiral perturbation theory; they describe
the process as an isospin conserving decay $\Dss\to\Ds\eta$ followed by
the conversion of the virtual $\eta$ meson into a \piz\ meson through
mixing. The radiative process  $\Dss\to\Ds\g$ proceeds via magnetic
transitions to which both the coupling to the charm quark and the light
quark contribute. Since a neutral pion or kaon do not couple to the
photon, only virtual states with a charged pion or kaon contribute to
one-loop corrections of the magnetic moments. The difference in the
kaon and pion masses naturally leads to $\mathrm{SU}(3)$
flavor-symmetry breaking.

A previous measurement of the ratio $\Gamma(\Dss\to\Ds\piz)/\Gamma(\Dss\to\Ds\g)$
by the CLEO Collaboration obtained a result of
$0.062\ppmm{0.020}{0.018}\stat\pm0.022\syst$ using $14.7\ppmm{4.6}{4.0}$
$\Dss\to\Ds\piz$ events~\cite{bib:CLEO}.

The analysis we present here, using $90.4\invfb$ of data recorded by the
\babar\ detector at the \pep2\ asymmetric \epem\ storage ring, represents a
significant improvement in precision and is particularly relevant given
the recent observations of two new \Ds\ meson
states~\cite{bib:Ds232,bib:Ds246}. The data have been recorded at and
approximately $40\mev$ below the $\FourS$ resonance.

%We have used $90.4\invfb$ of data recorded by the \babar\ detector
%at the \pep2\ asymmetric \epem\ storage rings operated at the Stanford Linear
%Accelerator Center. $9\gev$ electrons are collided with $3.1\gev$ positrons,
%resulting in a center-of-mass frame energy of $\sqrt{s} = 10.58\gev$.

The \babar\ detector is described in detail elsewhere~\cite{bib:NIM}.
Charged particles are detected and their momenta measured by a
silicon vertex tracker (SVT) consisting of five layers of double-sided
silicon strip sensors and a cylindrical 40-layer drift chamber (DCH), both
operating within a $1.5\mathrm{\,T}$ solenoidal magnetic field. 
Charged particle identification is provided by energy loss measurements
in the SVT and DCH and by light recorded in an internally reflecting ring 
imaging Cherenkov detector (DIRC).
Neutral particles are identified and their energies measured
by an electromagnetic
calorimeter (EMC) composed of 6580 CsI(Tl) crystals.

%A ring-imaging
%Cherenkov detector (DIRC) is used for identifying charged particles.
%Energies of neutral particles are measured by an electromagnetic calorimeter
%composed of $6\,580$ CsI(Tl) crystals.

\Ds\ mesons are reconstructed using the decay $\Ds\to\phi\pip$, $\phi\to\Kp\Km$.
Kaons are identified by combining the energy deposited in
the SVT and DCH with the DIRC information. Tracks not
identified as kaons according to the PID criteria are considered as pions.

Our selection criteria are chosen such that the signal significance $S^2/(S+B)$
of the channel $\Dss\to\Ds\piz$ is maximized. $S$ and $B$ refer to the expected
numbers of signal and background events, respectively, which are obtained from
a sample of Monte Carlo events.

All combinations of
$\Kp\Km\pip$ candidates have been required to successfully fit to a common
vertex; only combinations with a $\Kp\Km$ mass within $8\mevcc$ from the
nominal $\phi$ mass~\cite{bib:PDG2004} have been retained.

The scaled momentum is defined as
$x_{p}(\Ds) = p^{*}(\Ds) / p^{*}_{\mathrm{max}}(\Ds)$, where $p^{*}(\Ds)$ is
the momentum of the \Ds\ candidates in the center-of-mass frame, and
$p^{*}_{\mathrm{max}}(\Ds) = \sqrt{{E^{*}_{\mathrm{beam}}}^{2} - {m(\Ds)}^{2}}$
is its maximum value.
Combinations of $\phi$ and \pip\ candidates are retained if the scaled
momentum is $0.6$ or greater.
Since the kinematics of the \ccbar\ fragmentation process produce charmed
mesons with high momenta, this reduces the combinatorial background.

We exploit the longitudinal polarization of the $\phi$ meson
to reduce continuum background by requiring
that the absolute value of the cosine of the helicity angle $\theta_{H}$,
which is defined as the angle between the $\phi$ momentum direction in the
\Ds\ rest frame and the momentum direction of one of the kaons
in the $\phi$ rest frame, is $0.3$ or greater.

% An alternative
% method, which weights events with the second Legendre polynomial $P_{2}
% (\cos \theta_{H}) = 5/4 \left(3 \cos^{2} \theta_{H} - 1 \right)$, has been
% used as a cross-check and yields compatible branching ratios.

\begin{figure}
\includegraphics[width=\columnwidth]{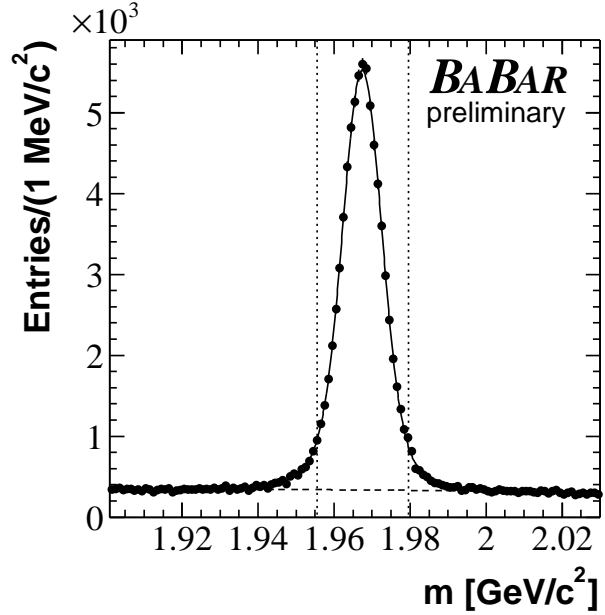}
\caption{\label{fig:Ds}$\Kp\Km\pip$ mass distribution.
The dots represent data points. The solid line shows the fitted function.
The dashed line indicates the portion of the fit associated with the
background. \Ds\ candidates are selected from the region between the
vertical dotted lines.}
\end{figure}

The $K^+K^-\pip$ invariant mass distribution is shown in Fig.~\ref{fig:Ds}.
By fitting the sum of a double Gaussian function (to represent the signal) and
a third-order polynomial (to represent the background) to this distribution,
we find $73\,500\pm300$ events. \Ds\ candidates are retained if their
invariant mass is within $12\mevcc$ from the nominal 
\Ds\ mass~\cite{bib:PDG2004}.

Each \piz\ candidate is reconstructed by combining two photon candidates.
Each photon candidate consists of a
calorimeter cluster which is not associated with a charged track and has an
energy in the laboratory frame of at least $45\mev$.
Additionally, to help remove the background from hadrons, we require 
the lateral moment~\cite{bib:LAT}, which describes the shape of the
electromagnetic shower in the calorimeter, to be less than $0.55$.

The \piz\ candidates are retained if they have a momentum $p^{*}$ in the
center-of-mass frame greater than $150\mevc$. Furthermore, the absolute
value of the cosine of the decay angle $\theta^{*}(\piz)$, which is
defined as the angle between the direction of one of the photons in the
\piz\ rest frame and the direction of the \piz\ candidate in the
center-of-mass frame, is required to be less than $0.85$. For the
isotropic \piz\ decay, the $|\cos \theta^{*}(\piz)|$ distribution is flat,
while it peaks near $1$ for random \gaga\ combinations.

Only \gaga\ pairs inside a mass window are retained. This window is
defined as the region where a function fitted to the \gaga\ mass
distribution of true $\piz\to\gaga$ Monte Carlo events exceeds $0.2$ its
maximum value, which accommodates the asymmetric shape of the \gaga\ mass
distribution and takes different detector calibrations into account. A
kinematic fit is applied to the surviving \gaga\ pairs, constrained to the
nominal \piz\ mass.

After combining the \Ds\ and \piz\ candidates to reconstruct the decay
$\Dss\to\Ds\piz$, we have fitted a function to the mass difference
$\Delta m (\Ds\piz) = m(\Kp\Km\pip\piz) - m(\Kp\Km\pip)$. The function
is the sum of a double Gaussian function to represent the signal and
\begin{eqnarray}
f_{1}(\Delta m) = N \cdot \left( 1 - \exp \left( - \frac{\Delta m - m(\piz)}{\mu} \right) \right)
                  \cdot \left( \Delta m^{2} + a\Delta m + b \right)
\label{eq:ExpQuad}
\end{eqnarray}
(where $m(\piz)$ is the \piz\ mass, while $N$, $\mu$, $a$, and $b$ are free
fit parameters) to represent the background. The exponential term in the
background function models the kinematic threshold; it is close to $1$ in
the signal region.

\begin{figure}
\includegraphics[width=\columnwidth]{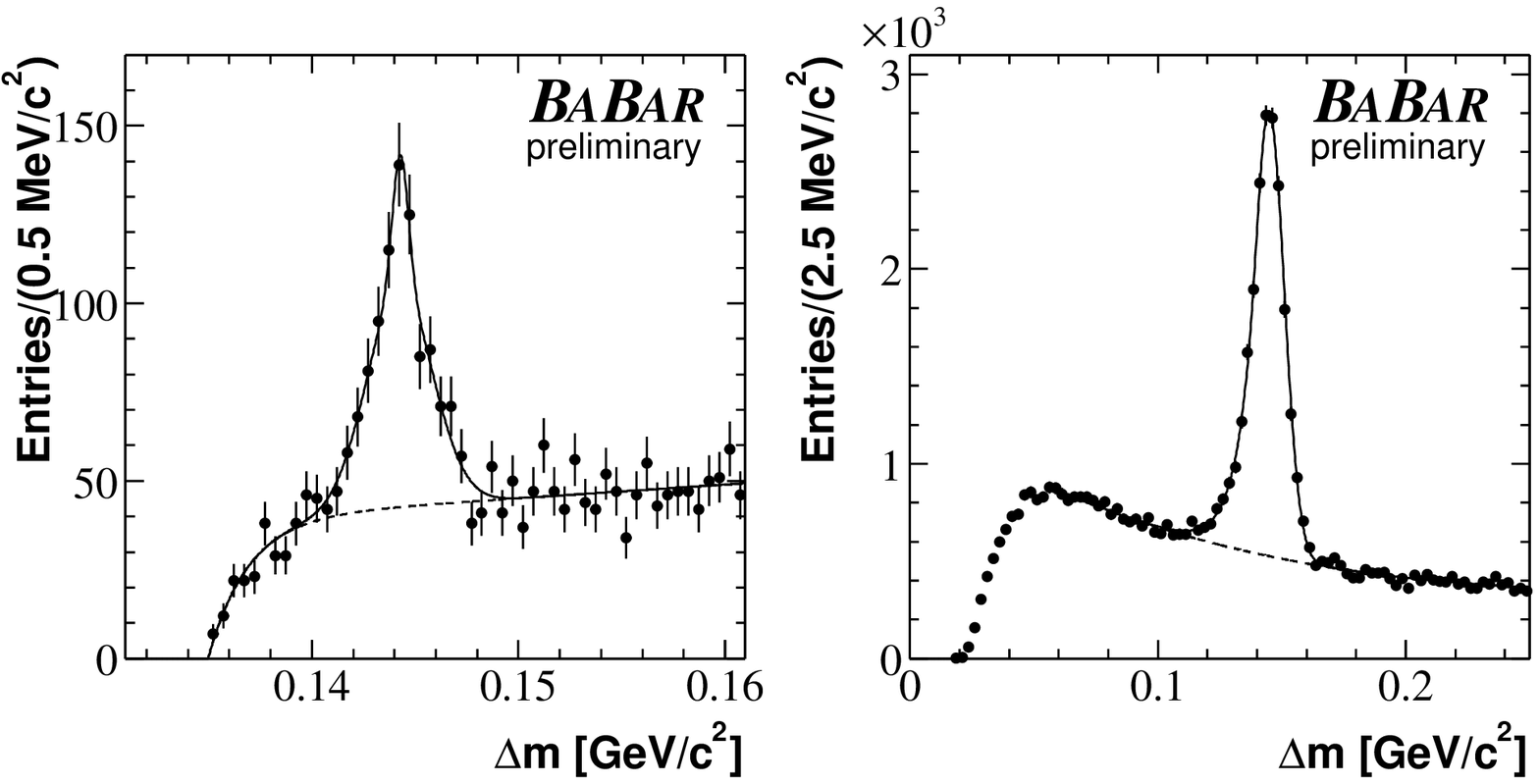}
\caption{\label{fig:Dss}\Dss\ signals:
(a)~$m(\Kp\Km\pip\piz)-m(\Kp\Km\pip)$;
(b)~$m(\Kp\Km\pip\g)-m(\Kp\Km\pip)$.
The dots represent data points. The solid line shows the fitted function.
The dashed line indicates the portion of the fit associated with the
background.}
\end{figure}

The result of this fit is shown in
Fig.~\ref{fig:Dss}(a). We find a signal event yield of $560\pm40$.

For the reconstruction of the decay $\Dss\to\Ds\g$, calorimeter clusters that
are not associated with a charged track are considered photon candidates if they
fulfill the following requirements. The energy must be $50\mev$ or greater in
the laboratory frame and $100\mev$ or greater in the center-of-mass frame. The
lateral moment must be less than $0.8$. To reduce the background of photons coming
from \piz\ decays, a photon candidate is discarded if it forms a \piz\ candidate
with any other photon candidate in the same event.
Here, a \gaga\ combination is
considered a \piz\ candidate if the invariant mass is in the range
$115 < m(\gaga) < 155\mevcc$ and has an energy of at least $200\mev$ in the
center-of-mass frame.

To obtain the $\Dss\to\Ds\g$ signal event yield, we have fitted a function 
to the mass difference $\Delta m (\Ds\g) = m(\Kp\Km\pip\g) - m(\Kp\Km\pip)$.
This function is a sum of a ``Crystal Ball'' function~\cite{bib:CrystalBall}
\begin{equation}
f_{2}(\Delta m) = N \cdot \left\{
  \begin{array}{ll}
  A \left( B - \frac{\Delta m-\mu}{\sigma} \right) ^{-n}         & {\mathrm{if}}\:(\Delta m-\mu)/\sigma \le \alpha \\
  \exp \left( - \frac {(\Delta m-\mu)^{2}}{2 \sigma^{2}} \right) & {\mathrm{if}}\:(\Delta m-\mu)/\sigma \ge \alpha
  \end{array} \right.
\label{eq:CrystalBall}
\end{equation}
(where $N$, $\mu$, $\sigma$, $n$, and $\alpha$ are free fit parameters,
and $A$ and $B$ are chosen such that the function and its first derivative
are continuous at $(\Delta m-\mu)/\sigma = \alpha$) for the signal, and a
third-order polynomial for the background. The fit result is shown
in Fig.~\ref{fig:Dss}(b). We find $15\,600\pm200$ signal events.

To determine the reconstruction efficiencies,
two samples of signal Monte Carlo events for the decays
$\Dss\to\Ds\piz$ and $\Dss\to\Ds\g$, each of which consists of
$30\,000$ events, have been used. The Monte Carlo events have
been analyzed using the same procedure as for real data. By
calculating the ratio of reconstructed to generated event numbers,
we find efficiencies of $\epsilon(\Ds\piz) = 0.041\pm0.002$
and $\epsilon(\Ds\g) = 0.071\pm0.002$ for the two \Dss\ decay
modes. The efficiency ratio is
$\epsilon(\Ds\piz)/\epsilon(\Ds\g) = 0.58\pm0.03$.

For completeness, and as a check of our efficiency calculations, 
we have also measured the ratio
$\Gamma(\Dstarz\to\Dz\piz)/\Gamma(\Dstarz\to\Dz\g)$ using the same
selection criteria for the \piz\ and photon candidates
as in the reconstruction of $\Dss\to\Ds\piz$ and $\Dss\to\Ds\g$. 
To reconstruct the decay $\Dz\to\Km\pip$, combinations of \Km\ and
\pip\ candidates are required to successfully fit to a common vertex, and
the scaled momentum of the resulting \Dz\ candidate must be $0.6$ or
greater. Fitting the sum of a double Gaussian function and a third-order
polynomial to the resulting $\Km\pip$ invariant mass distribution, we find
$(996.0\pm1.5)\times10^{3}$ signal events. $\Km\pip$ combinations are
retained if their mass differs by less than $17\mevcc$ from the nominal
\Dz\ mass~\cite{bib:PDG2004}.

\begin{figure}
\includegraphics[width=\columnwidth]{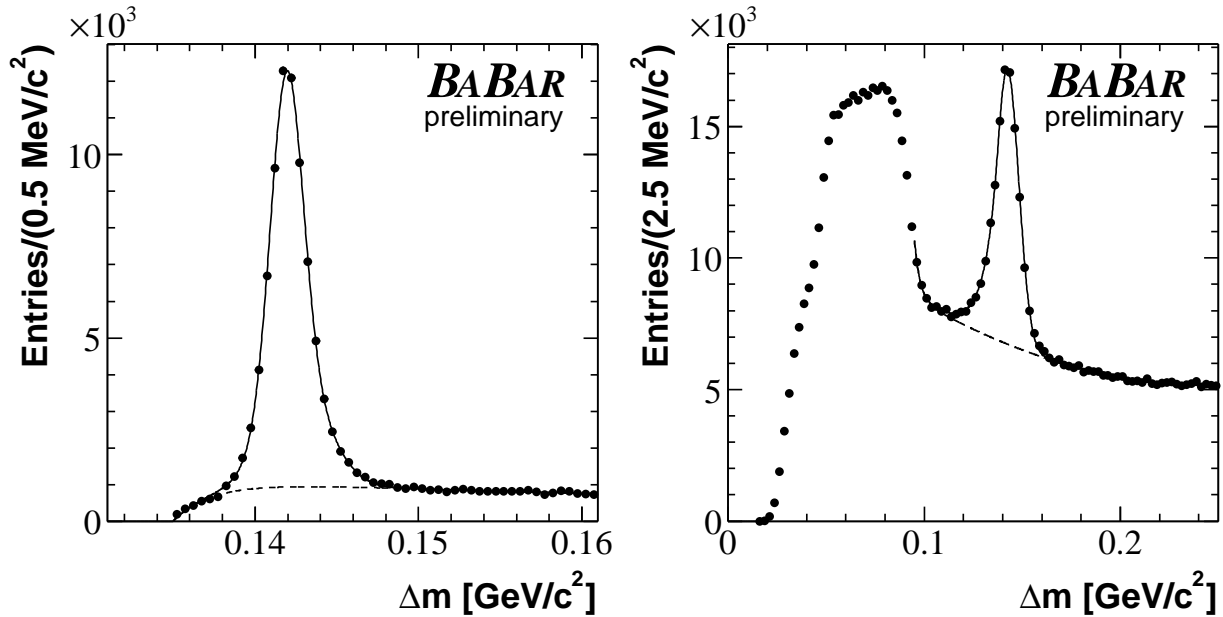}
\caption{\label{fig:Dstarz}\Dstarz\ signals:
(a)~$m(\Km\pip\piz)-m(\Km\pip)$;
(b)~$m(\Km\pip\g)-m(\Km\pip)$.
The dots represent data points. The solid line shows the fitted function.
The dashed line indicates the portion of the fit associated with the
background.}
\end{figure}

The \Dz\ candidates are combined with all \piz\ candidates; the resulting
mass difference $\Delta m (\Dz\piz) = m(\Km\pip\piz) - m(\Km\pip)$ is
shown in Fig.~\ref{fig:Dstarz}(a). A fit using a double Gaussian for the
signal and the function shown in Eq.~\ref{eq:ExpQuad} for the background
yields $69\,000\pm400$ signal events.

The \Dz\ candidates are then combined with all photon candidates producing
the distribution of the mass difference $\Delta m (\Dz\g) = m(\Km\pip\g) - m(\Km\pip)$
shown in Fig.~\ref{fig:Dstarz}(b). In this case the signal peak of
$\Dstarz\to\Dz\g$ is close to a large bump due to the reflection of $\Dstarz\to\Dz\piz$
where one photon comes from a \piz\ decay. We therefore model the background
using the function
\begin{eqnarray}
f_{3}(\Delta m) = N \cdot \left( 1 + \exp \left( - \frac{\Delta m - m(\piz)}{\mu} \right) \right)
                  \cdot \left( \Delta m^{2} + a\Delta m + b \right) .
\end{eqnarray}
(Note that this function is similar to Eq.~\ref{eq:ExpQuad},
but differs in the sign of the exponential term.)

The signal is modeled by a ``Crystal Ball'' function (Eq.~\ref{eq:CrystalBall}).
The resulting fitted signal consists of $67\,900\pm700$ events.

We determine the efficiency ratio by the reconstruction of
$30\,000$ signal Monte Carlo events for
each of the channels $\Dstarz\to\Dz\piz$ and $\Dstarz\to\Dz\g$.
Efficiencies of
$\epsilon(\Dz\piz) = 0.037\pm0.002$ and $\epsilon(\Dz\g) = 0.064\pm0.002$,
and an efficiency ratio of $\epsilon(\Dz\piz)/\epsilon(\Dz\g) = 0.58\pm0.03$
are found. This is in perfect agreement
with the value of $\epsilon(\Ds\piz)/\epsilon(\Ds\g) = 0.58\pm0.03$.
A ratio of $\Gamma(\Dstarz\to\Dz\piz)/\Gamma(\Dstarz\to\Dz\g) = 1.740\pm0.020$
(statistical error only) is obtained, which agrees with the world average of
$\Gamma(\Dstarz\to\Dz\piz)/\Gamma(\Dstarz\to\Dz\g) = 1.625\pm0.200$~\cite{bib:PDG2004}.

Various sources of systematic uncertainties have been studied; they are summarized
in Tables~\ref{tab:DssSystematics} and~\ref{tab:DstarzSystematics}.
% An uncertainty
% already mentioned is the statistical error arising in each fit to the signal
% Monte Carlo mass difference distributions.

\begin{table}
\caption{\label{tab:DssSystematics}Summary of systematic uncertainties on the
branching ratio\hglue 1.5in
\break$\Gamma(\Dss\to\Ds\piz)/\Gamma(\Dss\to\Ds\g)$.}
%\begin{ruledtabular}
\begin{center}
\begin{tabular}{lr@{.}l}
\hline
Signal Monte Carlo Statistics      &  5&0\% \\
Monte Carlo Simulation Uncertainty &  3&6\% \\
\Ds\  Background                   &  3&2\% \\
\piz\ Background                   &  3&1\% \\
Momentum Dependence                &  6&8\% \\ \hline
Total                              & 10&2\% \\
\hline
\end{tabular}
\end{center}
%\end{ruledtabular}
\end{table}

\begin{table}
\caption{\label{tab:DstarzSystematics}Summary of systematic uncertainties on the
branching ratio\hglue 1.5in
\break$\Gamma(\Dstarz\to\Dz\piz)/\Gamma(\Dstarz\to\Dz\g)$.}
%\begin{ruledtabular}
\begin{center}
\begin{tabular}{lr@{.}l}
\hline
Signal Monte Carlo Statistics      &  5&4\% \\
Monte Carlo Simulation Uncertainty &  3&8\% \\
\Dz\  Background                   &  0&1\% \\
Momentum Dependence                &  2&8\% \\ \hline
Total                              &  7&2\% \\
\hline
\end{tabular}
\end{center}
%\end{ruledtabular}
\end{table}

To verify that the Monte Carlo events model the data correctly, we have
studied \mtau\ decays with one or two \piz\ mesons in the final state to
obtain energy-dependent Monte Carlo efficiency correction functions for
\piz\ mesons and photons. While no correction is necessary, the errors on
the correction functions represent uncertainties of the Monte Carlo model
and are used as systematic uncertainties.

To test for uncertainties in the background shape of the mass difference
distributions, we have considered left and right sidebands in the \Ds,
\Dz, and \piz\ masses. The widths of the sidebands have been chosen such
that the number of events in each sideband corresponds to the expected
number of background events in the signal region. We have fitted the same
functions used to determine the signal yields to the mass difference
distributions of the sideband samples. Any discrepancy in the yield
between data and Monte Carlo simulation is considered a systematic
uncertainty.

We have repeated the measurements of $\Gamma(\Dss\to\Ds\piz)/\Gamma(\Dss\to\Ds\g)$
and $\Gamma(\Dstarz\to\Dz\piz)/\Gamma(\Dstarz\to\Dz\g)$
for candidates restricted to bins of $p^{*}$ (encompassing the range
$3 < p^{*} < 5\gevc$) by repeating the mass difference fit to the
subsample of candidates within each bin. By fitting both a constant function
and a first-order polynomial to each branching ratio as a
function of $p^{*}$, we have verified that the measured branching ratios
are independent of $p^{*}$.
Nevertheless, we assume conservatively that
the first-order polynomials arise from unknown momentum dependencies of the
efficiencies which do not cancel in the branching ratios. Taking the
momentum distributions of the \Dss\ and \Dstarz\ mesons into account, we
use the difference between the branching ratio represented by
the constant function and the first-order polynomial for each
distribution as a systematic uncertainty.

The branching ratios are shown in Table~\ref{tab:Results}.
By assuming that the \Dss\ meson decays only to $\Ds\piz$ and $\Ds\g$,
i.e.\ $\BR(\Dss\to\Ds\piz) + \BR(\Dss\to\Ds\g) = 1$, and that the
\Dstarz\ meson decays only to $\Dz\piz$ and $\Dz\g$,
i.e.\ $\BR(\Dstarz\to\Dz\piz) + \BR(\Dstarz\to\Dz\g) = 1$,
it is possible to calculate the branching fractions, which are also listed
in Table~\ref{tab:Results}.

In summary, we have measured a preliminary value for the branching ratio
$\Gamma(\Dss\to\Ds\piz)/\penalty-900\Gamma(\Dss\to\Ds\g) = 0.0621\pm0.0049\stat\pm0.0063\syst$,
which is consistent with the previous measurement~\cite{bib:CLEO}, but has higher
precision. We have also measured
$\Gamma(\Dstarz\to\Dz\piz)/\Gamma(\Dstarz\to\Dz\g) = 1.740\pm0.020\stat\pm0.125\syst$,
which is more precise than, and
in agreement with, the world average~\cite{bib:PDG2004}.

The branching fraction $\BR(\Dss\to\Ds\piz) = 0.0585\pm0.0043\stat\pm0.0056\syst$
is larger than the theoretical prediction, which is in the $1-3\%$ range~\cite{bib:ChoWise}.
However, it should be noted that the theory predicts a strong correlation between
$\BR(\Dss\to\Ds\piz)$ and $\BR(\Dstarp\to\Dp\g)$; for $\BR(\Dstarp\to\Dp\g) {\lsim} 1\%$,
$\BR(\Dss\to\Ds\piz)$ is expected to be greatly enhanced.

\begin{table}
\caption{\label{tab:Results}Summary of preliminary results. 
The first errors are statistical,
the second represent systematic uncertainties.}
%\begin{ruledtabular}
\begin{center}
\begin{tabular}{lc}
\hline
$\Gamma(\Dss\to\Ds\piz)/\Gamma(\Dss\to\Ds\g)$       & $0.0621\pm0.0049\pm0.0063$ \\
$\BR(\Dss\to\Ds\piz)$                               & $0.0585\pm0.0043\pm0.0056$ \\
$\BR(\Dss\to\Ds\g)$                                 & $0.9415\pm0.0043\pm0.0056$ \\ \hline
$\Gamma(\Dstarz\to\Dz\piz)/\Gamma(\Dstarz\to\Dz\g)$ & $1.740\dspace\pm0.020\dspace\pm0.125\dspace$  \\
$\BR(\Dstarz\to\Dz\piz)$                            & $0.6351\pm0.0027\pm0.0166$ \\
$\BR(\Dstarz\to\Dz\g)$                              & $0.3649\pm0.0027\pm0.0166$ \\
\hline
\end{tabular}
\end{center}
%\end{ruledtabular}
\end{table}

% Input the pub-board acknowledgements file
We are grateful for the excellent luminosity and machine conditions
provided by our \pep2\ colleagues, 
and for the substantial dedicated effort from
the computing organizations that support \babar.
The collaborating institutions wish to thank 
SLAC for its support and kind hospitality. 
This work is supported by
DOE
and NSF (USA),
NSERC (Canada),
IHEP (China),
CEA and
CNRS-IN2P3
(France),
BMBF and DFG
(Germany),
INFN (Italy),
FOM (The Netherlands),
NFR (Norway),
MIST (Russia), and
PPARC (United Kingdom). 
Individuals have received support from CONACyT (Mexico), A.~P.~Sloan Foundation, 
Research Corporation,
and Alexander von Humboldt Foundation.

\end{document}